\documentclass[preprint,12pt]{elsarticle}

\usepackage{graphics}
\usepackage{epsfig}

\usepackage{amssymb}

\usepackage{bm}
\def\beq{\begin{equation}}
\def\be{\begin{eqnarray}}
\def\eeq{\end{equation}}
\def\ee{\end{eqnarray}}

\def\bi{\bibitem}
\def\lsim{\buildrel < \over {_{\sim}}}
\def\gsim{\buildrel > \over {_{\sim}}}
\def\qm{{\bf q}}
\def\met{{1/2}}
\def\Veff{V_{{\rm eff}}}

\journal{Physics Letters B}

\begin{document}

\begin{frontmatter}



\title{Correlation effects on the weak response of nuclear matter}


\author[label1,label2]{Omar Benhar}
\author[label2,label1]{Nicola Farina}

\address[label1]{INFN, Sezione di Roma, I-00185 Roma, Italy}
\address[label2]{Dipartimento di Fisica, ``Sapienza'' Universit\`a di Roma, 
I-00185 Roma, Italy}

\begin{abstract}

The consistent description of the nuclear response at low
and high momentum transfer requires a unified dynamical model, suitable to
account for both short- and long-range correlation effects.
We report the results of a study of the charged current weak response of symmetric nuclear
matter, carried out using an effective interaction obtained from a realistic model
of the nucleon-nucleon force within the formalism of correlated basis functions.
Our approach allows for a clear identification of the kinematical regions in which
different interaction effects dominate.

\end{abstract}

\begin{keyword}

nuclear matter \sep neutrino scattering \sep effective interaction


21.65.-f \sep 24.10.Cn \sep 25.30.Pt


\end{keyword}

\end{frontmatter}



\section{Introduction}
\label{intro}

Short-range nucleon-nucleon correlations have long been recognized to play a
prominent role in the electromagnetic response of nuclei. 
Comparison between the longitudinal
responses at momentum transfer $|{\bf q}|~\gsim $~1.5 fm$^{-1}$, measured by inclusive 
electron scattering experiments, and the predictions of
independent particle models shows that the data in the region of the
quasi-free peak are sizably overestimated \cite{rl1,rl2}. On the other hand,
the results of approaches based on Nuclear Many-Body Theory (NMBT) and realistic
models of the NN interaction \cite{fp,ff}, in which the effects of short-range correlations
are taken into account, provide a satisfactory description
of a large body of data \cite{jourdan}.

Long-range correlations, leading to the excitation of collective modes, are
also known to be important. They become in fact dominant
in the region of low momentum transfer, in which the space resolution of the probe,
$\lambda \sim |{\bf q}|^{-1}$, becomes much larger than the average NN separation.
Most theoretical studies of this kinematical regime have been carried out within the
Tamm Dancoff (TD) \cite{TD} and Random Phase Approximation (RPA) \cite{GPC},
using effective particle-hole interactions.

The development of a theoretical scheme allowing for a consistent treatment of
short- and long-range correlations requires the construction of {\em effective} 
interactions based on {\em realistic} models of the NN force.
This goal has been pursued by the authors of Ref.\cite{shannon1}, who used the
formalism of correlated basis functions (CBF) and the cluster expansion
technique \cite{feenberg,CBF1} to obtain an
effective interaction from a truncated version of the state-of-the-art Argonne $v_{18}$
potential \cite{argonne}.
The resulting effective interaction and the associated effective transition
operators have been applied in a systematic study of the weak response of 
nuclear matter \cite{shannon1}.

The development of a unified description of the nuclear response to weak
interactions at low and high energy is important, its quantitative
understanding being needed in a variety of different contexts. The region
of neutrino energy $E_\nu~\sim~0.5~-~3$~GeV \cite{PRD} is relevant
to the analyses of long-baseline neutrino experiments, such as K2K and MiniBOONE
(see, e.g., Ref.\cite{NUINT} and references therein), while nuclear interactions
of low energy neutrinos, carrying energies of the order of tens of MeV, are believed
to determine supernov\ae$\ $evolution \cite{burrows} and neutron star cooling \cite{yako}.

The effective interaction proposed in Ref.\cite{shannon1} has been recently improved
with the inclusion of the effects of three- and many-nucleon forces \cite{valli}.
The resulting interaction has been used to calculate the nucleon-nucleon scattering
cross section in nuclear matter, needed to obtain the transport coefficients
from the Boltzmann-Landau equation \cite{valli,ictp}.

In this paper, we report the results of a study of the charged current weak
response of symmetric nuclear matter, carried out using the effective interaction
of Ref.\cite{valli}. 
The main purpose of our work is identifying different interaction and
correlation effects, all described within the same dynamical model, and 
determining the kinematical regions in which they play the dominant role.

\section{Formalism}
\label{formalism}

In the kinematical regime in which the non relativistic approximation
is applicable, the nuclear response to a weak probe delivering momentum ${\bf q}$ and
energy $\omega$ can be written in the form
\beq
S(\qm,\omega)=\frac{1}{N}\sum_n \ \langle 0|O^\dagger({\qm})|n\rangle\langle
n|O({\qm})|0\rangle\delta(\omega+E_0-E_n) \ .
\label{ris:def}
\eeq
where $N$ is the particle number and $|0\rangle$ and $|n\rangle$ denote the initial and 
final nuclear states, respectively.
 The transition operators are obtained expanding the weak nuclear current in powers
of $|{\bf q}|/m$, $m$ being the nucleon mass.
In the case of charged current interactions, at leading order one finds the Fermi and
Gamow-Teller operators, whose expressions in coordinate space are
($R \equiv ({\bf r}_1,\ldots,{\bf r}_{N}$))
\beq
\langle R^\prime | O^F(\qm) | R \rangle = \delta( R-R^\prime ) \ 
 \ g_V \sum_{i=1}^N \ {\rm e}^{i\qm{\bf r}_i} \tau_i^+ \ 
 =  \delta( R-R^\prime ) \ \sum_{i=1}^N \  O^F_i(\qm) \ ,
\label{def:opf}
\eeq
\beq
\langle R^\prime | O^{GT}(\qm) | R \rangle = \delta( R-R^\prime ) \ 
g_A \sum_{i=1}^N \ {\rm e}^{i\qm{\bf r}_i} \bm{\sigma}_i\tau_i^+ \ =
 \delta( R-R^\prime ) \   \sum_{i=1}^N \  O^{GT}_i(\qm) \ .
\label{def:opgt}
\eeq
In the above equations, ${\bf r}_i$ specifies the position of the $i$-th particle, 
$\bm{\sigma}_i$ describes its spin and $\tau_i^+$ is the isospin rising operator. 

In our formalism nuclear matter is described using correlated states, obtained 
from the Fermi (FG) gas states through the 
transformation \cite{feenberg,CBF1}
\begin{equation}
| n \rangle = \frac{ F | n ) }
{ ( n | F^\dagger F | n )^{1/2} } \ ,
\label{def:corr_states}
\end{equation}
where $| n )$ is a determinant of single particle states
representing $N$ noninteracting nucleons at density $\rho = 2 k_F^3/3 \pi^2$, $k_F$ being the 
Fermi momentum. The operator $F$, embodying the correlation structure 
induced by the non perturbative components of the NN interaction, is written in the form
\begin{equation}
F(1,\ldots,N)=\mathcal{S}\prod_{j>i=1}^N f_{ij} \  ,
\label{def:F}
\end{equation}
where $\mathcal{S}$ is the symmetrization operator, accounting for the fact that, in general,
$\left[ f_{ij} , f_{ik} \right] \neq 0$.

The two-body correlation function $f_{ij}$ features an operatorial structure reflecting
the complexity of the NN potential:
\beq
f_{ij} = \sum_{TS} \left[ f_{TS}(r_{ij}) + \delta_{S1} f_{tT}(r_{ij}) S_{ij} \right] \ P_{TS} \ .
\eeq 
In the above equation $r_{ij} = |{\bf r}_i - {\bf r}_j|$, $P_{TS}$ is the operator projecting onto 
two-nucleon states of total spin and isospin $S$ and $T$, respectively, and 
$S_{ij}~=~3({\bm \sigma}_i~\cdot~{\bf r}_{ij})
({\bm \sigma}_j \cdot {\bf r}_{ij})/r_{ij}^2~-~({\bm \sigma}_i \cdot {\bm \sigma}_j)$. 
The shapes of the radial functions $f_{TS}(r_{ij})$ and $f_{tT}(r_{ij})$, are determined
through functional minimization of the expectation value of the nuclear hamiltonian in
the correlated ground state, carried out at the two-body level of the cluster 
expansion \cite{cfcns}.

The CBF formalism naturally leads to the appearance of an effective interaction, acting on the FG 
basis states. At lowest order of CBF perturbation theory $\Veff$ is {\em defined} 
by
\beq
\label{veff:1}
\langle H \rangle = \frac{ ( 0 |F^\dagger H F| 0 ) }{ ( 0 | F^\dagger F| 0 )} 
=  \frac{ ( 0 |F^\dagger (T + V)  F| 0 ) }{ ( 0 | F^\dagger F| 0 )} =
( 0 | T + \Veff | 0 ) \ ,
\eeq
where $H = T + V$ is the nuclear hamiltonian, $T$ is the kinetic energy operator 
and $V~=~\sum_{j>i} v_{ij}$, $v_{ij}$ being the NN potential.

As the above equation suggests, in principle the approach based on the effective interaction 
allows one to obtain any nuclear matter observables using perturbation theory in the 
FG basis. However, the calculation of the expectation value of the hamiltonian 
 in the correlated ground state, needed to extract $V_{{\rm eff}}$ from
Eq.(\ref{veff:1}), involves severe difficulties. 

In this work we follow the procedure developed in Ref. \cite{shannon1}, whose 
authors derived the expectation value of $\Veff$ carrying out a 
cluster expansion of $\langle H \rangle$ of Eq.(\ref{veff:1}), and 
keeping only the two-body cluster contribution. The resulting expression is
\beq
 ( 0 | \Veff | 0 )
 = \sum_{j < i} \langle ij | f_{12} \left[ -\frac{1}{m} (\nabla^2 f_{12}) + 
v_{12} f_{12} \right] | ij \rangle_a \ ,
\label{veff:2}
\eeq
where the laplacian operates on the relative coordinate and the subscript $a$ 
indicates that the two-nucleon state $|ij\rangle_a$ is antisimmetryzed.

We have computed the effective potential defined by the above equation using the 
truncated form of the Argonne $v_{18}$ potential referred to 
as $v^\prime_{8}$ \cite{v8p}. Three-nucleon forces, whose inclusion is 
needed to reproduce the saturation properties of nuclear matter, 
have been described following the approach of Ref. \cite{LagPan}, in 
which the main effect of three- and many-body interactions is taken into
account through a density dependent modification of $v_{ij}$ at 
intermediate range.
The Euler Lagrange equations for the correlation functions $f_{TS}$ and $f_{tT}$
have been solved using correlation ranges taken from Ref. \cite{APR1}.

The calculation of transition matrix elements between correlated states is also 
non trivial. While in the FG model the Fermi and Gamow-Teller operators can only induce
transitions to one particle-one hole (1p1h) states, in the presence of correlations more
complex final states yield nonvanishing contributions to the response. In addition, the
numerical determination of the matrix elements requires the use of the cluster expansion 
technique.

In this work we only include transitions to {\em correlated} 1p1h states and approximate 
the transition matrix element according to
\beq
M_{ph} = \frac{ ( ph | F^\dagger O F | 0 )}
{ ( ph |F^\dagger F| ph )^\met ( 0|F^\dagger F|0)^\met} 
\approx ( ph|(1+\sum_{j>i}g_{ij}) O (1+\sum_{j>i}g_{ij})|0 ) \ ,
\label{Mph}
\eeq
where $g_{ij} = f_{ij} -1$ and $O=\sum_i O_i$, $O_i$ being the Fermi or Gamow-Teller operator 
(see Eqs.(\ref{def:opf}) and (\ref{def:opgt})).

The authors of Ref.\cite{shannon1} used a different truncation scheme to obtain $M_{ph}$ from the 
cluster expansion formalism. Our choice is motivated by the requirement of consistency with the 
definition of the effective interaction, Eq.(\ref{veff:2}). However, we have 
verified that the difference between the numerical results obtained from the two different 
prescriptions never exceeds few percent.

\section{Effects of short-range correlations}
\label{src}

The calculation of $M_{ph}$ has been performed on a cubic lattice, with a discrete set of 
$N_h$ states specified by the hole momentum ${\bf h_i}$, satisfying the conditions 
${\bf h_i}~<~k_F$ and ${\bf h_i}~+~{\bf q}~>~k_F $.
 The size of the basis has been determined requiring that the response of a system of noniteracting 
nucleon computed on the lattice agreed with the analytical result of the FG model.
Obviously, the response obtained from the discrete set of final states consists of a collection of
delta function peaks. A smooth function of $\omega$ has been obtained using a gaussian
representation of the energy conserving $\delta$-function of finite width $\sigma$. 
For sufficiently small values of $\sigma$, the results become independent of $\sigma$.

\begin{figure}[ht]
\centerline{\includegraphics[scale=0.45]{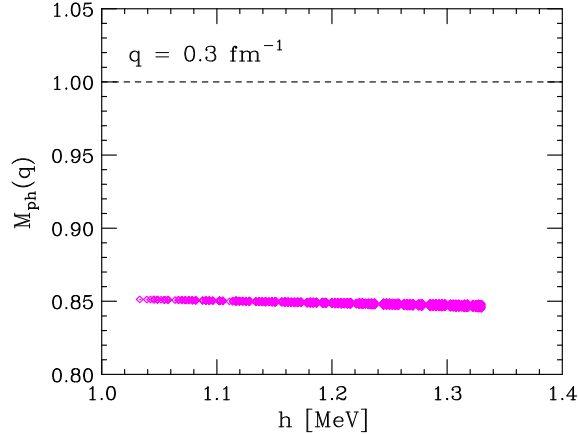}}
\caption{\small Fermi transition matrix element at $|{\bf q}|=0.3 \ {\rm fm}^{-1}$ 
(Eq.(\ref{Mph})) as a function of the magnitude of hole momentum $|{\bf h}|$. The 
calculation has been 
carried out using a basis of 3040 ph states. The dashed horizontal
line corresponds to the result of the FG model.  \label{quenching}}
\end{figure}

Figure \ref{quenching} shows a comparison between correlated and FG matrix elements for the 
case of Fermi transitions. Note that the effect of the sizable quenching produced by short-range 
correlations is enhanced in the calculation of the response, whose definition involves the 
correlated matrix element squared (see Eq.(\ref{ris:def})). Similar results are obtained for the 
Gamow-Teller transitions.

The calculation of the response requires the knowledge of the energies of the initial 
and final states. In the case of 1p1h final states, the argument of the energy conserving 
$\delta$-function appearing in Eq.(\ref{ris:def}), reduces to $E_n - E_0 = e_{\bf p} - e_{\bf h}$, 
$e_{\bf k}$ being the energy
of a nucleon of momentum ${\bf k}$ in nuclear matter. Within our approach, $e_{\bf k}$ can be
calculated using the effective interaction and the Hartree-Fock (HF) approximation: 
\beq
e_{\bf k}=\frac{{\bf k}^2}{2m}+\sum_{|{\bf h}| < k_F} \langle h k|\Veff|h k\rangle_a  
 = \frac{{\bf k}^2}{2m}  + U_{HF}({\bf k}) \ .
\label{singlespectrum}
\eeq
The numerical results turn out to be in close agrement with those of Ref. \cite{bob:ep}, 
obtained within the CBF approach using the Fermi-Hyper-Netted-Chain (FHNC) summation 
technique and a realistic nuclear hamiltonian.

\begin{figure}[ht]
\centerline{\includegraphics[scale=0.85]{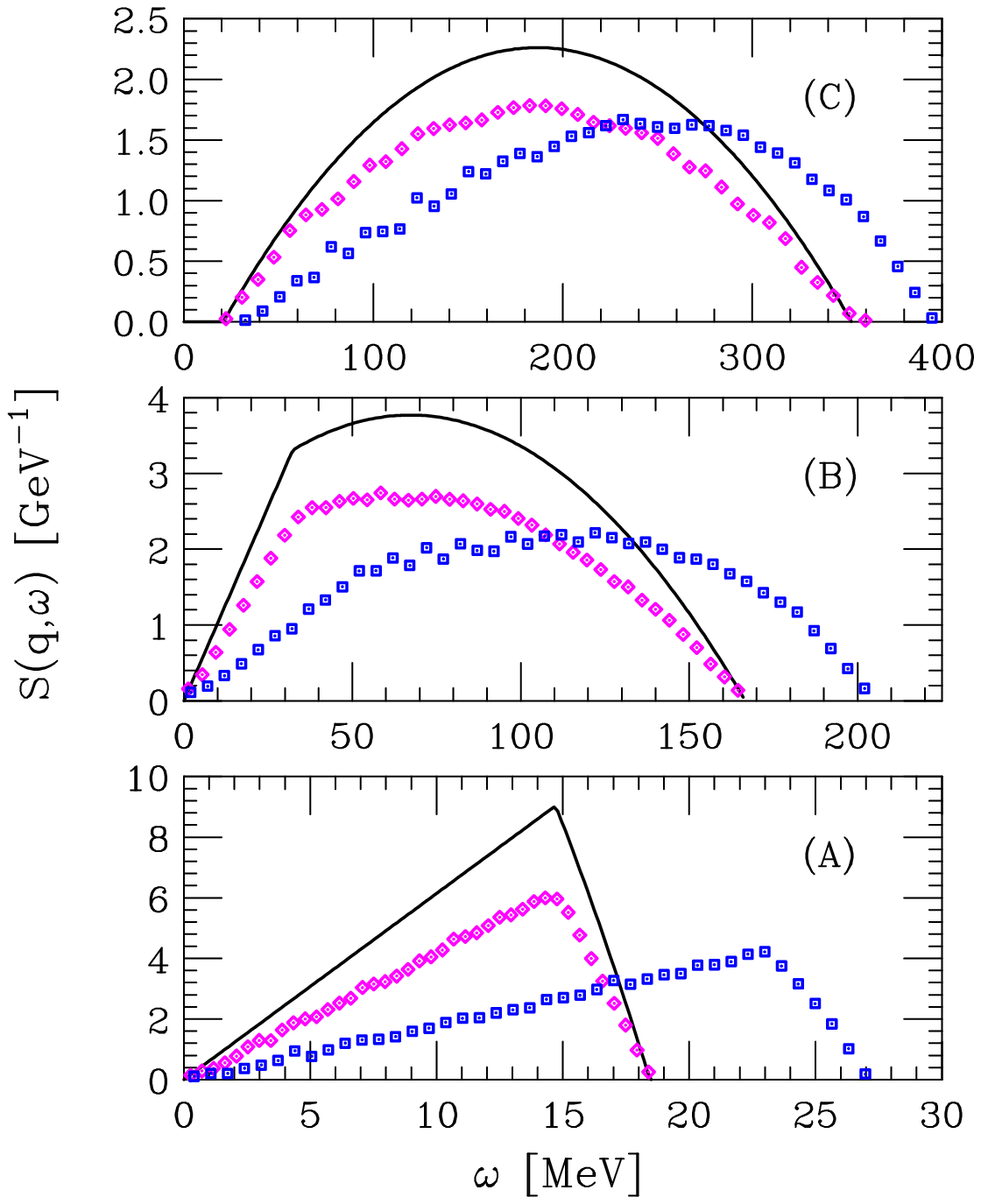}}
\caption{\small Nuclear matter response for the case of Fermi transitions.
Solid lines: FG model. Squares: correlated HF approximation. Diamonds: same as the
squares, but with the HF spectrum of Eq.(\ref{singlespectrum}) replaced by the
kinetic energy spectrum.
Panels (A), (B) and (C) correspond to $|{\bf q}|=$0.3, 1.8 and 3.0 fm$^{-1}$, respectively.
  \label{HF}}
\end{figure}

Figure \ref{HF} shows the nuclear matter response for the case of Fermi transitions. 
Panels (A), (B) and (C) correspond to the 
kinematical regions $|{\bf q}|<k_F$, $k_F < |{\bf q}| < 2k_F$ and $|{\bf q}|>2k_F$, respectively. 
The solid lines represent the results of the FG model, while the squares show the response
obtained within the correlated HF approximation, i.e. using the correlated matrix elements of 
Eq.(\ref{Mph}) and the energy spectrum of Eq.(\ref{singlespectrum}). In order to clearly identify 
the $\sim 30$ \% quenching due to short-range correlations, whose size is consistent with the 
results of Fig. \ref{quenching}, we also show by diamonds the response obtained replacing the HF 
spectrum of Eq.(\ref{singlespectrum}) with the kinetic energy spectrum, i.e. neglecting 
$U_{HF}({\bf k})$ in Eq.(\ref{singlespectrum}). It appears that interaction effects 
leading to the appearance of the mean field also play a significant role, pushing strength 
to energies well beyond the kinematical limit of the FG model.
Panels (A), (B) and (C) correspond to momentum transfer $|{\bf q}|=$0.3, 1.8 and 3.0 fm$^{-1}$
and basis size $N_h=$3040, 2637 and 2777, respectively.
The same qualitative pattern is obtained in the case of Gamow-Teller transitions.

\section{Effects of long-range correlations}
\label{lrc}

Equation (\ref{Mph}) shows that in the correlated HF approximation the Fermi and
Gamow-Teller transition operators are replaced by effective operators, acting on FG states, 
{\em defined} through
\beq
( ph |(1+\sum_{j>i}g_{ij}) O (1+\sum_{j>i}g_{ij})|0 ) = 
 ( ph |O_{{\rm eff}}|0 ) \ .
\label{def:Oeff}
\eeq
Note that the above effective operators are different from those of Ref.\cite{shannon1}, 
as we have chosen a different truncation scheme of the cluster expansions.

The FG ph states, while being eigenstates of the HF hamiltonian
\beq
H_{HF} = \sum_{{\bf k}} e_{\bf k} \ ,
\label{def:HHF}
\eeq
with $e_{\bf k}$ given by Eq.(\ref{singlespectrum}), are not eigenstates of the full 
nuclear hamiltonian. As a consequence, there is a residual interaction $V_{{\rm res}}$
that can induce transitions between different ph states, as long as 
their total momentum, {\bf q}, spin and isospin are conserved. 

We have included the effects of these transitions, using the Tamm Dancoff (TD) approximation,
 which amounts to expanding the final state in the basis of
one 1p1h states according to \cite{Boffi}
\beq
| f ) = | {\bf q}, \ T S M ) = \sum_{i} c^{TSM}_{i}
| p_i h_i, \ T S M ) \ ,
\label{tda:phexp}
\eeq
where ${\bf p}_i = {\bf h}_i + {\bf q}$, $S$ and $T$ denote the total spin and isospin of 
the particle-hole pair and $M$ is the spin projection. 

At fixed ${\bf q}$, the excitation energy of the state $| f )$, $\omega_{f}$, as well 
as the coefficients $ c^{TSM}_{i}$, are determined solving the eigenvalue equation 
\beq
H | f ) = (H_{HF} + V_{{\rm res}})| f ) = (E_0 + \omega_{f}) | f ) \ ,
\label{eigenprob}
\eeq
where $E_0$ is the ground state energy. Within our approach this amounts to diagonalizing a 
$N_h \times N_h$ matrix whose elements are 
\beq
H^{TSM}_{ij} = (E_0 + e_{{\bf p}_i} - e_{{\bf h}_i}) \delta_{ij}
 + ( h_i p_i, \ T S M | V_{{\rm eff}} | h_j p_j, \ T S M ) \ ,
\label{Hmat1}
\eeq

In TD approximation, the response can be written as
\beq
S(\qm,\omega)=  \sum_{TSM}  \ \sum_{n=1}^{N_h} 
\left| \sum_{i=1}^{N_h}  (c^{TSM}_n)_i 
( h_i p_i, \ T S M | O_{{\rm eff}}(\qm)|0 ) \right|^2  
\delta(\omega-\omega^{TSM}_n) \ ,
\label{tda:resp}
\eeq
where $(c^{TSM}_n)_i$ denotes the $i$-th component of the eigenvector belonging to the 
eigenvalue $\omega^{TSM}_n$. 

\begin{figure}[ht]
\begin{center}
\includegraphics[scale=0.85]{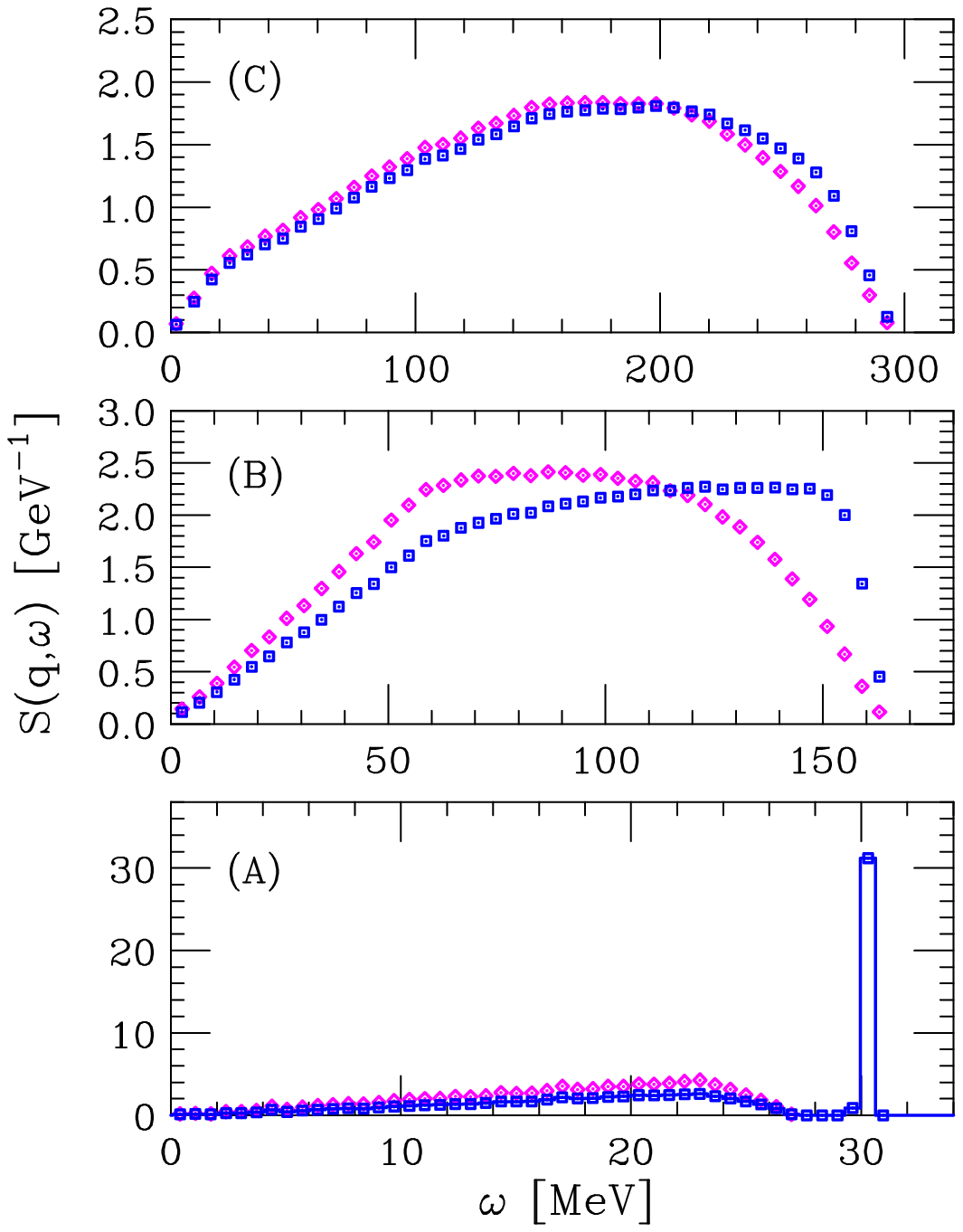}
\end{center}
\caption{\small Nuclear matter response calculated within the TD (squares) and
correlated HF (diamonds) approximations, for the case of Fermi transitions.
Panels (A), (B) and (C) correspond to $|{\bf q}|=$0.3 , 1.5 and 2.4 fm$^{-1}$,
respectively.  \label{TD} }
\end{figure}

The diagonalization has been performed using a basis of 
$N_h\sim 3000$ ph states for each spin-isospin channel. 
The appearance of an eigenvalue lying outside the particle hole continuum, corresponding 
to a collective excitation reminiscent to the plasmon mode of the electron gas, is clearly 
visible in panel (A) of Fig. \ref{TD}, showing the TD response at $|{\bf q}|=$ 0.3 fm$^{-1}$ 
for the case Fermi transitions. For comparison, the result of the correlated HF approximation 
is also displayed. Note that the sharp peak arises from the contributions of particle-hole 
pairs with $S=1$ $T=0$. 

In order to identify the kinematical regime in which long range correlations 
 are important, we have studied
 the TD response in the region 0.3 $ \leq |{\bf q}| \leq$ 3.0 fm$^{-1}$.
The results show that at $|{\bf q}| \lsim $1.2 fm$^{-1}$ the peak correponding to the 
collective mode in the $S=1$ $T=0$ channel is still visible, althoug less prominent. 
However, it disappears if the exchange contribution to the matrix element of the 
effective interaction appearing in the rhs of Eq.(\ref{Hmat1}) is neglected.

The transition to the regime in which short-range correlations dominate is
illustrated in panels (B) and (C) of Fig. \ref{TD}, showing the comparison between 
TD and HF responses at $|{\bf q}|=$ 1.5 and 2.4 fm$^{-1}$, respectively.

At $|{\bf q}|=$1.5 fm$^{-1}$ the peak no longer 
sticks out, but the effect of the mixing of ph states with $S=1$ and $T=0$ is still 
detectable, resulting in a significant enhancement of the strength at large $\omega$. 
At $|{\bf q}|=$2.4 fm$^{-1}$ the role of long range correlations
turns out to be negligible, and the TD and correlated HF responses come very close to 
one another. The calculation of the response associated 
with Gamow-Teller transitions shows a similar pattern.

\section{Conclusions}
\label{summary}

The CBF formalism employed in our work is ideally suited to construct an effective 
interaction starting from a realistic NN potential. The resulting effective interaction, 
which has been shown to provide a quite reasonable account of the equation of state
of cold nuclear matter \cite{valli}, allows for a consistent description of 
the weak response in the regions of both low and high momentum transfer, where 
different interaction effects are important.
 
The results of our calculations suggest that in addition to the HF mean field, which  
moves the kinematical limit of the transitions to 1p1h states well beyond the FG value,
correlation effects play a major role, and must be taken into account.
While at $|{\bf q}| \lsim 0.5$ fm$^{-1}$ long-range correlations, leading to the appearance 
of a collective mode outside the particle-hole continuum, dominate, at 
$|{\bf q}| \gsim 2.0$ fm$^{-1}$ the most prominent effect is the quenching due to 
short-range correlations. 

As a final remark, we emphasize that the results discussed in this paper, while being 
certainly interesting in their own right, should be seen as a step towards the
development of a unified description of a variety of nuclear matter properties, 
 relevant to the understanding of neutron-star structure and dynamics \cite{valli,ictp}.



\end{document}